\documentclass[twocolumn,prp,aps]{revtex4}
\usepackage{amsmath}
\usepackage{amssymb}
\usepackage{graphicx}
\usepackage{dcolumn}
\usepackage{graphicx}
\usepackage{dcolumn}
\usepackage{bm}

\begin{document}


\title{Robust helical edge transport at $\nu=0$ quantum Hall state}
\author{G. M. Gusev,$^1$ D. A. Kozlov,$^{2,3}$ A. D. Levin,$^1$ Z. D. Kvon,$^{2,3}$ N. N. Mikhailov,$^2$ and S. A. Dvoretsky,$^{2}$}

\affiliation{$^1$Instituto de F\'{\i}sica da Universidade de S\~ao
Paulo, 135960-170, S\~ao Paulo, SP, Brazil}

\affiliation{$^2$Institute of Semiconductor Physics, Novosibirsk
630090, Russia}

\affiliation{$^3$Novosibirsk State University, Novosibirsk 630090,
Russia}

\date{\today}
\begin{abstract}
Among the most interesting predictions in two-dimensional materials with a Dirac cone is the existence of the zeroth Landau level (LL),
equally filled by electrons and holes with opposite chirality. The gapless edge states with helical spin structure emerge from Zeeman splitting
 at the LL filling factor $\nu=0$  gapped quantum Hall state.
 We present observations of a giant nonlocal four-terminal transport in zero-gap HgTe quantum wells at the $\nu=0$ quantum Hall state.
 Our experiment clearly demonstrates the existence of the robust helical edge state in a system with single valley Dirac cone materials.

\end{abstract}
\maketitle
Two dimensional massless Dirac fermions in the presence of a strong perpendicular magnetic field show several remarkable features that sharply diverge from conventional
behaviour [1-8]. The energy spectrum is organized in Landau levels (LL) with square root versus linear dependence on the magnetic field and square root dependence on the Landau index $n$ versus $n+1/2$, in comparison
with  the parabolic dispersion at the zero field.  The most remarkable consequence of this last property is the existence of a zero-energy Landau level ($n = 0$).
This is not due to the linear spectrum, but is related to the $\pi$ Berry phase carried by each Dirac point. Therefore, the $n = 0$ LL has a magnetic field independent energy,
which is quite different from a quantized cyclotron orbit in the conventional quantum Hall effect.
It is important to find the clear experimental signature that can help to identify zeroth Landau level in many Dirac materials,
such as graphene [2], three-dimensional topological insulators [9] and Weyl semimetals [10].

The existence of the zeroth Landau level has been examined by measurements of the integer quantum Hall effect (QHE) in graphene [2]. Previous experiments in samples with moderate mobility provide indirect evidence of the QHE around filling factor $\nu=0$: the Hall conductivity has a peculiar plateau at
$\sigma_{xy} = 0$ that is not precisely quantized as other plateaus are, and its longitudinal resistivity does not vanish [11].
But this interpretation is based on the bulk spectrum scenario (figure 1a).
Note, however, that understanding of the QHE at $\nu=0$ requires
a presence of edge states similar to the conventional integer QHE. In this case, the interpretation becomes ambiguous and depends on the particular structural properties of graphene.
In particular, one of the scenarios predicts that, if spin splitting is larger than valley splitting, the bulk Landau level forms two counter propagating edge states [12]
similar to 2D topological insulators [13] as shown in Figure 1b.
It is worth noting that unambiguous experimental support for the existence of counter-propagating edge states
is provided by nonlocal measurements. The helical edge state transport at  $\nu=0$ differs from the chiral edge mode transport for a higher LL:
chiral states carry the same chemical potential in the vicinity of each boundary, while counter circulating edge states carry potential from different current probes (left and right).
As a result, conductance is zero in the QHE regime and quantized in universal units $2e^{2}/h$ in the QH-metal regime in the absence of backscattering between spin-polarized states.
Several attempts have been made to study nonlocal transport in graphene, however, opposing or conflicting interpretations have been offered [14,15].
Very recently observation of the quantized local and nonlocal resistances in a single layer [16] and in bilayer [17] graphene
of micrometer-sized samples in the presence of the strong in-plane magnetic field has been reported, which has been attributed to the parallel B-induced helical edge modes.
Application of other materials that posses a single Dirac cone is of particular interest.

\begin{figure}[ht]
\includegraphics[width=7cm]{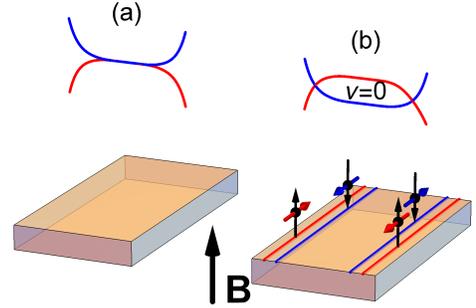}
\caption{\label{fig.1}(Color online) Schematics of band structure (energy spectrum) in low (a) and high (b) magnetic field, showing the zero
LL in the middle of the sample and at the sample edge, and counter propagating spin polarized edge states in a slab-shaped sample for the $\nu=0$ LL state. }
\end{figure}

Recently a two-dimensional system with a single Dirac cone spectrum, based on HgTe quantum wells, has been discovered  [18,19].
The single spin degenerate Dirac valley allows unambiguous identification of the features resulting from the bulk zeroth Landau level. In addition, the high mobility and giant Lande g-factor ($\sim 55$)
favor formation of spin-polarized counter propagating states. In the present paper, we studied the
nonlocal transport in 10-probe devices fabricated from HgTe zero-gap quantum structures.
We observe a magnetic field induced, giant, nonlocal resistance peak near the CNP in different configurations of current and voltage probes. The nonlocal response is comparable with
local resistance and increases rapidly with B. The nonlocal resistance persists in magnetic fields up to 7 T. Simple Kirchhoff based estimations and
more complicated edge state+bulk model calculations clearly confirm the existence of helical edge states originating from the bulk zeroth LL.

\begin{figure}[ht]
\includegraphics[width=8cm]{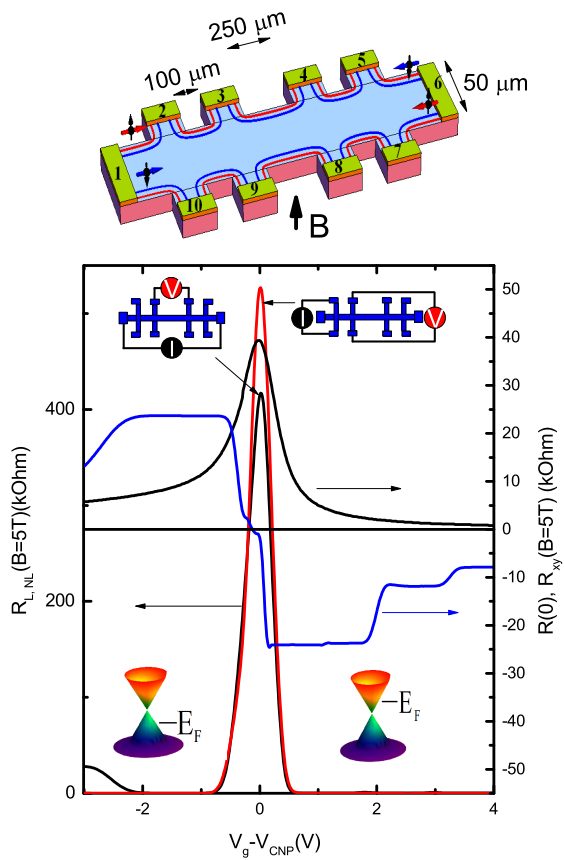}
\caption{\label{fig.2}(Color online) Top-a schematic of device (not preserving aspect ratio) and the local zero field resistance R(0),
longitudinal  $R_{L}=R_{xx}$ (I=1,6, V=3,4)(black curve) Hall $R_{xy}$ (I=1,6; V=3,9)(blue curve)
and nonlocal $R_{NL}$ (I=2,10; V=3,9) (red curve) resistances as a function of gate voltage at B=5 T, T=4.2 K, $I=10^{-9}$A.
The schematics show how the current source and the voltmeter are connected for the
measurements.}
\end{figure}

Quantum wells $Cd_{0.65}Hg_{0.35}Te/HgTe/Cd_{0.65}Hg_{0.35}Te$ with (013) surface orientations and a well thickness of 6.3, 6.4 and 6.6 nm were prepared by molecular beam epitaxy.
A detailed description of the sample structure has been given in Refs. [17,18].  The sample is a Hall bar device with 8 voltage probes. The bar has a width $W$ of $50 \mu m$
and three consecutive segments of different lengths $L$ $(100, 250, 100 \mu m )$ (fig.2). A dielectric layer was deposited (100 nm of $SiO_{2}$ and 100 nm of $Si_{3}Ni_{4}$) on the sample surface and
then covered by a TiAu gate. The density variation with gate voltage was $1\times10^{11} cm^{-2}V^{-1}$. The magnetotransport measurements were performed in the temperature range
$1.4 - 70 K$  using a standard four point circuit with a $1-13 Hz$ ac current of $1-10 nA$ through the sample,
which is sufficiently low to avoid overheating effects. It is worth noting that the electrical contacts to the electron gas beneath the gate electrode become worse in
a strong magnetic field, therefore we report the results up to 5 or 7 T (depending on the particular device).  10 devices from three different wafers were measured, all with similar results.

    Figure 2 shows the zero field, longitudinal $R_{xx}$, Hall $R_{xy}$  and nonlocal $R_{NL}$ resistances
measured in a perpendicular magnetic field B=5 T as a function of gate voltage. $R_{xx}$ and $R_{xy}$ are measured in multiterminal Hall bar geometry.
In the local configuration, the current flows between contacts 1,6 and voltage is measured between probes 3,4 ($R_{L}=R_{xx}=R_{1-6,3-4}=V_{3,4}/I_{1,6}$) and
Hall voltage is measured between probes 3,9 ($R_{xy}=R^{3,9}_{1,6}=V_{3,9}/I_{1,6}$). In the nonlocal configuration,
the current flows between contacts 2,10 and voltage is measured between probes 3,9 ($R_{NL}=R^{3,9}_{2,10}=V_{3,9}/I_{2,10}$).
Zero field resistance behaviour resembles behaviour in other HgTe-based quantum wells, including topological insulators [13,19,20] and semimetals [23,24]:
resistance shows a peak around the charge neutrality point (CNP). In graphene and zero gap HgTe wells, the CNP is coincident with the Dirac point [5,7,18,19]. The maximum resistivity
at the Dirac point $\rho_{xx}=\frac{W}{L}R^{3,4}_{1,6}(0)=0.3\frac{h}{e^{2}}$ agrees with others' observations [18,20].

\begin{figure}[ht]
\includegraphics[width=10cm]{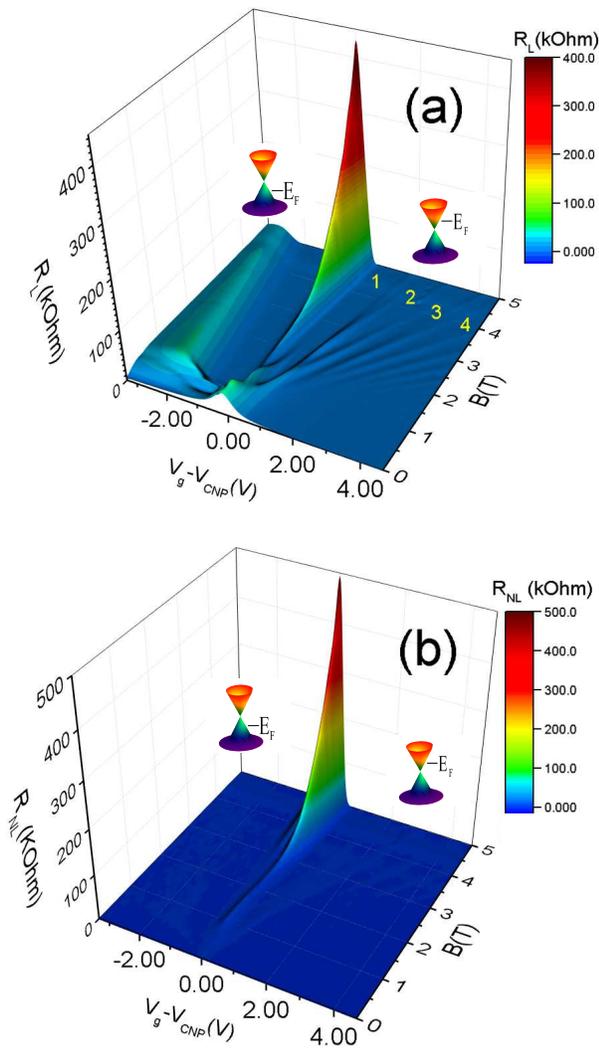}
 \caption{\label{fig.3}(Color online) The local $R_{L}=R_{xx}$ (I=1,6, V=3,4) (a)
and nonlocal $R_{NL}$ (I=2,10; V=3,9) (b) resistances as a function of the gate voltage and magnetic field, T=4.2K. Filling factors determined from Hall resistance are labeled.}
\end{figure}

When we applied an external perpendicular magnetic field a pronounced anomaly in the resistance data was observed:
resistance was found to increase very strongly with B near the CNP, while in other regions the system demonstrates conventional quantum Hall behaviour (fig.2).
Evolution of the local $R_{L}=R^{3,4}_{1,6}=V_{3,4}/I_{1,6}$ and nonlocal $R_{NL}=R^{3,9}_{2,10}=V_{3,9}/I_{2,10}$ resistances with gate voltage and magnetic field is shown in Figure 3.
Both $R_{L}$ and $R_{NL}$ exhibit sharp peaks above the critical magnetic field $B_{c} \sim 2.5 T$. Nonlocality is absent in the
magnetic field below $B_{c}$.

\begin{figure}[ht]
\includegraphics[width=8cm]{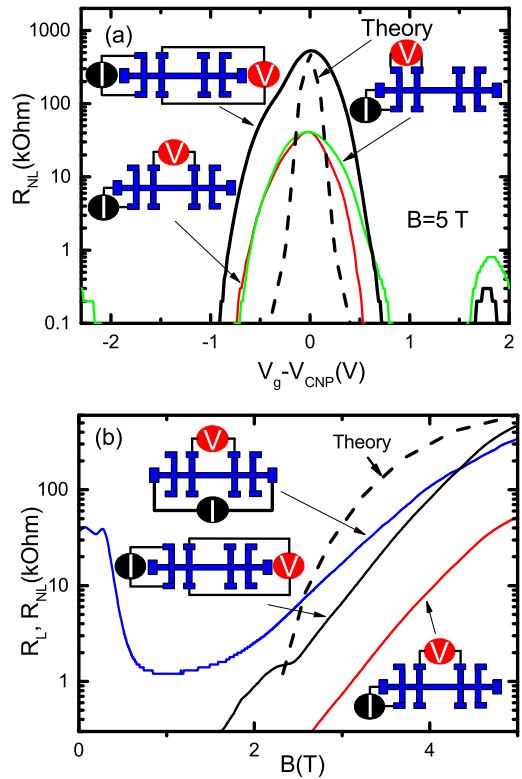}
\caption{\label{fig.4}(Color online)(a) The nonlocal $R_{NL}$ resistances as a function of gate voltage at B=5 T, T=4.2 K, obtained
for different measuring configurations.
Dashes - nonlocal resistance $R_{NL}$ (I=2,10; V=3,9) calculated from model [23]. (b)
Comparison of the magnetic field dependencies for both local and nonlocal resistances in various configurations obtained at the CNP.
Dashes - nonlocal resistance $R_{NL}$ (I=2,10; V=3,9) calculated from model [23]. The schematics show how the current source and the voltmeter are connected for the
measurements.}
\end{figure}

The important difference between the quantum Hall states with $\nu=0$ and $\nu\neq 0$ is that, in the conventional quantization regime, the
longitudinal transport coefficient vanishes in both conductivity $\sigma_{xx}$ and resistivity $\rho_{xx}$, while for the QHE state in $\nu=0$ this is not necessarily the case $\rho_{xx}=0$.
Indeed we obtain $\sigma _{xy}\sim \sigma_{xx}\sim 0$ because $\rho_{xx}\gg \rho_{xy}$ at $\nu=0$ [23]. Note that for the Hall insulating state
it is expected that both longitudinal and Hall resistivities are going to infinity near the CNP in accordance with the classical Hall resistivity formula
$\rho_{xy}\sim B/(n-p)ec$, where n and p are electron and hole densities respectively, if we assume the same mobility for both carrier types. Divergent longitudinal
and vanishing Hall resistivities have been observed at the Dirac point in graphene and attributed to density inhomogeneities
associated with electron-hole puddles [26].  The alternative approach to the $\nu=0$ quantum Hall effect was based on counter-propagating
edge channels with opposite spin directions [12]. In this model the Hall resistance is zero because of compensation between the helical states in accordance with Landauer-Buttiker formalism,
while the resistance measured between probes is quantized in units of $h/2e^{2}$ in the ballistic case and much higher than the resistance quantum in the diffusive case, similar
to a 2D topological insulator in zero magnetic field [22,27].

We provide new evidence that advances this debate
by measuring the long range nonlocal transport in the quantum Hall effect regime in Dirac cone materials. For example, because graphene has two valleys, two different situation must be considered, depending
on whether the bulk valley splitting is larger or smaller than the bulk spin splitting in the zeroth LL. If the valley
degeneracy lifts first, and valley separation becomes larger than the Zeeman energy,
the edge states do not cross and the gap appears both in the bulk and at the edges of the sample. Such a quantum Hall insulator resembles
common insulators without edge states. The other possibility occurs when the Zeeman energy is larger than the
valley splitting. The electron-like (hole-like) Landau level bends upwards (downwards) in energy near the
edge of the sample and forms two counter-propagating edge states residing on the same edge (figure 1b). Such QH-metal resembles a nontrivial
topological insulator in a zero magnetic field with a bulk gap and helical edge states protected by
the time reversal symmetry against backscattering [13]. The mechanism of the lift of fourfold degeneracy and the existence
of a possible insulating state in graphene have been discussed extensively [28-32]. A high-field insulating state
has been observed in both low and high quality graphene samples. The general consensus is that the magnetic field drives graphene at the CNP from the
QH-metal state at a relative low field into the QH-insulator state at a high enough field.  Note, however, the presence of
electron-hole puddling due to inhomogeneity can mask metallic behaviour at low magnetic fields and observation of the B-independent resistance peak
 may not be sufficient for the actual manifestation of the zeroth LL.

Figure 4a illustrates $R_{NL}$ for various configuration measured at a fixed magnetic field B=5T. Figure 4b shows the magnetic field dependence
of both local and nolocal resistances at fixed gate voltage corresponding to the CNP. One can see almost exponential growth of $R_{L}$ and $R_{NL}$ above $B_{c}$.
In the magnetic field region $0.3 T < B < 1. 4 T$, the experimental local resistance shows complex and diverse behaviour:
it reveals a large negative magnetoresistance and plateaux -like features. Previously [20] these features have been attributed to the emergence
of the zeroth Landau level without edge states (see figure 1a), which is represented by a sharp peak near the CNP in $\sigma_{xx}(V_{g})$.
When the magnetic field is such that the bulk Zeeman gap is larger than the zeroth LL broadening $\Gamma$, which occurs at $B_{c}\approx \Gamma/g\mu_{B}\approx 2.5 T$,
where $\mu_{B}$ is the Bohr magneton, one expects nonlocal transport due to the helical edge states.
The nonlocal resistance is strongly suppressed at high temperatures, while the local resistance is found to be much more robust [27].

To get more insight into the physics of the observed nonlocality, it is important to estimate the phenomenon theoretically.
To explain the value of the local and nonlocal resistances in most of the experimental observations
mentioned above, a simple picture of the helical edge state is usually enough. We have seen that, using Kirchhoff’s laws,
we can analyze our circuit. There is a simple expression that allows one to calculate the resistance
value for any measurement configuration assuming that there is only edge state transport in the sample [27]:
$R_{n,m}^{i,j}=\frac{L_{n,m}L_{i,j}}{Ll} (h/e^2)$,
where $R_{n,m}^{i,j}$ is the voltage measured between contacts $i$ and $j$ while the current is maintained between contacts $n$ and
$m$, $L_{i,j}$ ($L_{n,m}$) are the distances between $i$ and $j$ ($n$ and $m$) along the gated sample edge that does not include $n$ and
$m$ ($i$ and $j$), $L$ is the total perimeter of the sample, and $l$ is the mean free path due to the scattering between helical states
propagated along the same edge. Assuming a homogeneous material and that the mean path remains constant along the edge,
we obtain the ratio between nonlocal and local resistances: $R_{NL}/R_{L}=R^{3,9}_{2,10}/R^{3,4}_{1,6}=2$;
$R_{NL}/R_{L}=R^{2,3}_{1,10}/R^{3,4}_{1,6}=0.11$; $R_{NL}/R_{L}=R^{3,4}_{1,10}/R^{3,4}_{1,6}=0.125$.
The result of these calculations is close to the experimental result: $(R_{NL}/R_{L})^{exp}=R^{3,9}_{2,10}/R^{3,4}_{1,6}=1.1-1.75$;
$(R_{NL}/R_{L})^{exp}=R^{2,3}_{1,10}/R^{3,4}_{1,6}=0.08-0.1$; $(R_{NL}/R_{L})^{exp}=R^{3,4}_{1,10}/R^{3,4}_{1,6}=0.07-0.1$.
There is an interesting fact that, in certain sample probe configurations, one can obtain a nonlocal resistance greater than the local one (figs.2,3).

However, away from the CNP, the current penetrates into the interior of the sample, and the system behaves more and more
like a conventional two-dimensional gas. We apply formalism developed in Ref.10 to describe the local resistance in graphene near $\nu=0$ in QH metal
and then extended it to nonlocal resistance in semimetals [33]. The model explains the transport coefficients
in the regime, where the edge-state current is suppressed by bulk contribution to conductivity.
The scattering between the edge states and the scattering between bulk states and each of the edge states is characterized by the mean free path
$\gamma^{-1}=l$ and $g^{-1}$ respectively, which are assumed to be smaller than the sample's dimensions [25].

Figs.4 a,b display the results obtained from this model for typical parameter values $\gamma^{-1}= 0.33 \mu m$ and $g^{-1}= 0.33 \mu m$.
Clearly, the model reproduces qualitatively the main features of the measurements, in particular, suppression of the peak away from the CNP by
bulk contribution to the transport and rapid growth with the magnetic field due to Zeeman splitting. Note that calculated peak profile
is sharper than the measured one. A detailed comparison with the experiment requires knowledge of the density of the states in the gap between the Landau levels, and
ratio between localized and delocalized electrons on the tails of the Gaussian density of states in the quantum Hall regime.

A number of the theoretical models have addressed the evolution of the helical edge states in a magnetic field [34-36]. The results appear to be controversial:
while primary models predict that the counter propagating edge states persist up to a critical magnetic field [34,35], more recent calculations
demonstrate the emergence of a gap in the spectrum of the edge states at an arbitrary small B [36]. Such high sensitivity of the edge state spectrum
to the external magnetic field has been attributed to the natural interface inversion asymmetry in HgTe quantum wells. In sharp contrast with this prediction, we
observe a giant nonlocal magnetoresistance, which confirms the persistence of the helical states up to 7 T.  Note that the effective Hamiltonian
which ascribes the bulk energy spectrum and the structure of the edge states in the presence of the perpendicular magnetic field is not properly derived
from microscopic theory. For example, other $6\times 6$ matrix Hamiltonian has been successfully applied to the calculation of the
energy spectrum in HgTe quantum wells in the presence of an in-plane magnetic field [37]. Further theoretical study is required.

In conclusion, we have studied the local and nonlocal transport properties of the zero Landau level in a Dirac cone 2D system based on a zero-gap HgTe quantum well.
A giant, nonlocal resistance has been observed at the $\nu=0$ quantum Hall effect.
In comparison with graphene, which is the most typical 2D material with Dirac cones, our system has several advantages.
Firstly, the single cone spectrum allows unambiguous interpretation of the integer Hall effect based on the existence of counterpropagating edge states with opposite spin
emerging  from the zero bulk LL, ruling out the valley first split model in graphene. Secondly, the advantage in fabrication (MBE growth versus exfoliation) allows the production of samples with a large distance between probes and, therefore, justifies the long range nature of the nonlocal transport due to the edge states.

 The financial support of this work by the Russian Science Foundation (Grant No. 16-12-10041, MBE growth of HgTe QWs, fabrication of the field effect transistors and carrying out of the experiment and data analysis), FAPESP (Brazil) and CNPq (Brazil) is acknowledged. The authors thank O. E. Raichev for helpful discussions.
 
\section{Online supplemental material to : Robust helical edge transport at $\nu=0$ quantum Hall state}

\subsection{Resistivity and conductivity tensors and $R_{xx}(B,n_{s})$ diagram}
In the main text, we demonstrate the magnetoresistance and Hall effect as a function of the gate voltage at a fixed magnetic field.
Longitudinal resistance reveals a giant peak near the CNP with values $R_{xx}>>h/e^{2}$, and the Hall effect demonstrates bipolar behaviour.
Sometimes it is instructive to convert the resistivities to a conductivity tensor through a matrix inversion.
The quantum Hall state with $\nu=0$ distinguishes states with $\nu\neq 0$. For example, in the conventional quantization regime,
the longitudinal transport coefficient vanishes in both conductivity $\sigma_{xx}$ and resistivity $\rho_{xx}$, while for the QHE state in $\nu=0$
the longitudinal resistance shows maxima rather than minima and $\rho_{xx}\gg \rho_{xy}$.
\begin{figure}[ht]
\includegraphics[width=8cm]{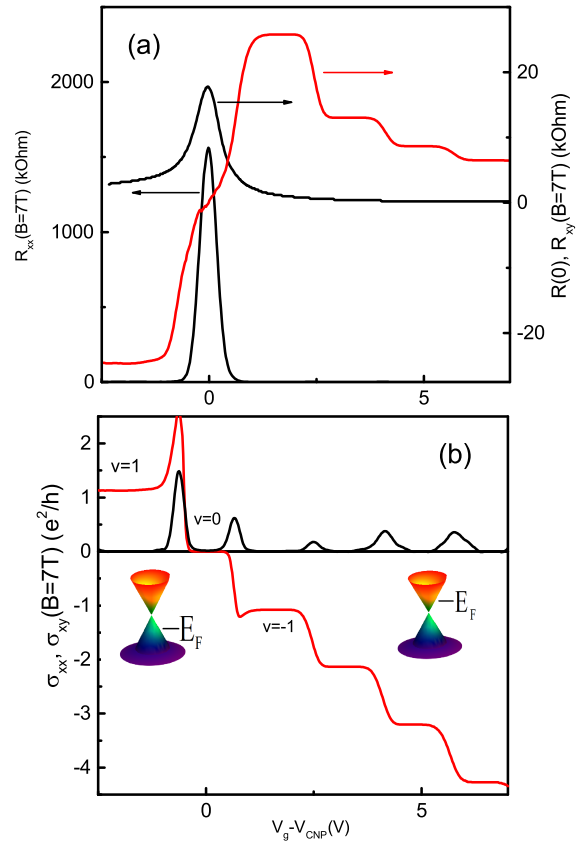}
\caption{\label{fig.1}(Color online) (a)  The local zero field resistance R(0), longitudinal  $R_{L}=R_{xx}$ (I=1,6, V=2,3)(black curve) and  Hall $R_{xy}$ (I=1,6; V=3,9)(red curve)
 resistances as a function of gate voltage at B=7 T, T=4.2 K, $I=10^{-9}$A.
 (b) The longitudinal and Hall conductivities as a function of gate voltage at B=7T.}
\end{figure}
Figure 5a shows transport coefficients in zero magnetic field and at B=7 T as a function of gate voltage.
Indeed we obtain $\sigma _{xy}\sim \sigma_{xx}\sim 0$ because $\rho_{xx}\gg \rho_{xy}$ at $\nu=0$, as is shown in Figure 5b.
Most importantly, however, this does not prove the existence of the Hall insulating state,
and further measurements of the nonlocal transport is required. The results of such measurements are presented in the main text.
Because the resistance peak at the CNP is larger than the magnetoresistance oscillations related to higher Landau levels, it is convenient
to visualize the quantum Hall effect through a $R_{xx}(B, n_{s})$ diagram.

Figure 6 displays the evolution of longitudinal resistance with a magnetic field and density ($R_{xx}(B, n_{s}$). One can see stripes corresponding to resistance maxima and minima in the $B-n_{s}$ plane for electron-like states.
It is worth noting that there is no direct correspondence between the experimental $R_{xx}(B, n_{s}$ plot and the LL spectrum because of the oscillating behaviour of Fermi energy.
The Dirac LL spectrum has a square root dependence on the magnetic field, while the experimental $R_{xx}(B, n_{s}$  diagram shows a linear $n_{s}$ versus B dependence,
where slopes of the stripes are determined by the LL filling factors $\nu$ : $\frac{dB}{dn}=\frac{\nu e}{h}$. Filling factors determined from the slopes are coincident with those determined from the Hall resistance.
\begin{figure}[ht]
\includegraphics[width=9cm]{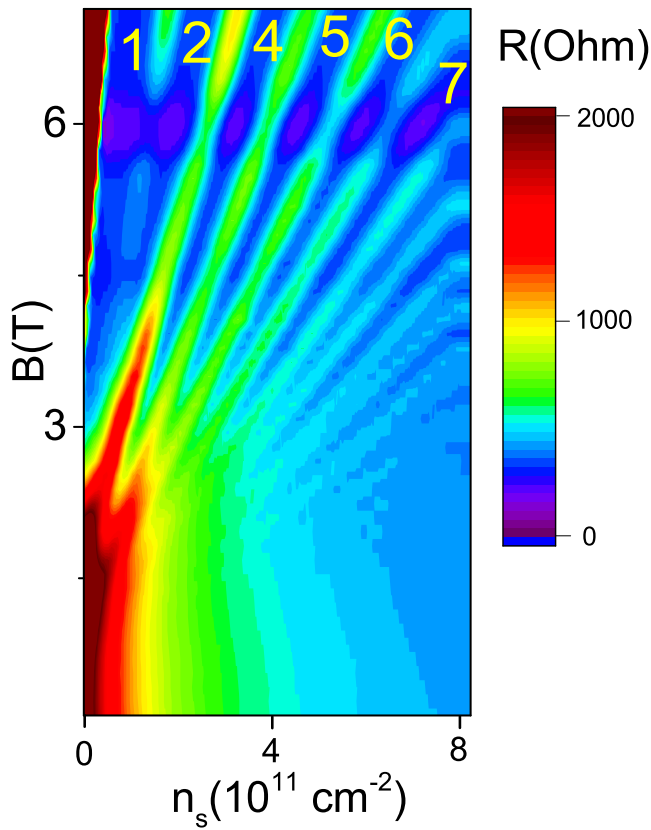}
\caption{\label{fig.2}(Color online) (a) $R_{xx}(B, n_{s})$ diagram, T=4.2 K. Filling factors $\mid\nu\mid$ determined from Hall resistance are labeled.}
\end{figure}

\subsection{Edge and bulk transport model and comparison with experiment}

 The mechanism responsible for the observed sharp peak of the local and nonlocal resistivities near the charge neutrality point (CNP) in zero gap HgTe quantum wells relies on the combination of the edge state and bulk transport contributions
with account of the backscattering within one edge as well as
bulk-edge coupling. When the gate voltage is swept through the CNP,
the local and nonlocal transport coefficients arise from the edge
state contribution at the CNP and the short-circuiting of the edge transport
by the bulk contribution away from the CNP.

Our analysis of transport coefficients in the main text was based on
a simplified model developed for the explanation of the quantum Hall effect
state in graphene at $\nu=0$, when the spin degeneracy is lifting before valley splitting [12].
Within this scenario, the transport
in graphene in a strong magnetic field is dominated by a pair of gapless counter propagating chiral
edge modes with opposite spin.  The transport properties in the bulk can be
described by the current-potential relation

\begin{eqnarray}
\textbf{j}_{i} =-\hat{\sigma}_{i}\nabla\psi_{i},
\end{eqnarray}
\[
\hat{\sigma}_{i} = \left( {\begin{array}{cc}
 \sigma_{xx}^{(i)} & \sigma_{xy}^{(i)} \\
 -\sigma_{xy}^{(i)} & \sigma_{xx}^{(i)}  \\
 \end{array} } \right), i=1,2
\]

where $\psi_{1,2}$ are the electrochemical potentials for electron-like and
hole-like states. So we can solve the problem by solving the Laplace equation
for these potentials $\nabla^{2}\psi_{1,2}=0$ because of the charge
conservation continuity conditions $\nabla\textbf{j}_{i}=0$. A
solution to Laplace's equation is uniquely determined if the value
of the function is specified on all boundaries. In order to describe
the transport properties in the presence of the edge states, we
introduce two phenomenological constants $\gamma$ and $g$, which
represent edge to edge and bulk to edge inverse scattering length
respectively. The boundary conditions describing the bulk edge
coupling are given by
\begin{eqnarray}
\textbf{n}\textbf{j}_{i} =g(\psi_{i}-\varphi_{i}),
\end{eqnarray}

where $\textbf{n}$ is a normal vector to the boundary, $\varphi_{i}$
are local chemical potentials of the electron-like and hole-like
edge states.

The edge state transport can be described by the equations for particle
density [12], taking into account the
scattering between the edge and the bulk

\begin{eqnarray}
\partial_{x}\varphi_{1}=\gamma(\varphi_{2}-\varphi_{1})+g(\psi_{1}-\varphi_{1}),\\
-\partial_{x}\varphi_{2}=\gamma(\varphi_{1}-\varphi_{2})+g(\psi_{2}-\varphi_{2}),
\end{eqnarray}

Another couple of similar equations describe the variables
$\varphi_{1^{'}}$ and  $\varphi_{2^{'}}$ at the opposite edge.

The general solution of this problem, therefore, includes the solution
of the 2D Laplace equation for the bulk electrochemical potentials
$\psi_{1,2}$ together with 8 equations (2,3,4) describing the
scattering between electron-like and hole-like edge states and
between the edge and bulk states. The current can be calculated from
this solution as a sum of the contribution from the bulk and
both edge states.
 We assumed that the bulk conductivities for electrons
and holes are represented by Gaussians in $e^{2}/h$ units
\begin{eqnarray}
\sigma_{xx}^{(1)} =e^{-(\frac{E-\mu}{\Gamma})^2}, \sigma_{xx}^{(2)} =e^{-(\frac{E+\mu}{\Gamma})^2},
\end{eqnarray}
where the parameter $\Gamma$ is the width of the Landau level (LL), $E=\frac{g\mu_{B}B}{2}$ is Zeeman splitting,
$\mu$ is the electrochemical potential.  In the
Gaussian  models, the  Landau level  width  is  proportional  to $B^{1/2}$.
Because Zeeman valley splitting is not well known, we can rewrite equations (5) in terms of the LL filling factor $\nu$ [1]:
\begin{eqnarray}
\sigma_{xx}^{(1,2)} =e^{-A(\nu\pm 1)^2}
\end{eqnarray}

The Hall conductivities are determined from the semicircle relation
\begin{eqnarray}
\sigma_{xy}^{(1,2)}(\sigma_{xy}^{(1,2)}\mp 2)+
(\sigma_{xx}^{(1,2)})^{2}=0
\end{eqnarray}
The full edge $+$ bulk model in the local resistance configuration
can be solved either analytically or numerically. The analytical
solution has been obtained in Ref.12 for graphene. It is worth noting that the model describes the local resistance behaviour as a function of energy. A detailed comparison
with the experiment requires knowledge of the density of the states in the gap between the Landau levels because the Fermi energy does not vary linearly within the bulk gap region. In
the absence of disorder, the Fermi level jumps from the
electron-like LL to the hole-like LL, and a sharp resistance
peak is expected, in contrast to the broad maximum
observed in the experiment. The existence of
metallic puddles could be responsible for a smoother
Fermi level displacement, similar to graphene. For simplicity's sake,
we can assume that the fraction of the puddles between the LL is constant, which leads to a constant
density of states inside the bulk gap $D_{0}$. Comparing the energy and the density scales,
we obtain $D_{0}=0.8x10^{10} cm^{-2}meV^{-1}$. Electrons in the puddles are localized and do not contribute to conductivity.

The edge $+$ bulk model
in the nonlocal resistance configuration can only be solved by numerical
methods. We have performed a self consistent calculation to find the $\psi
_{1,2}$ solution of the Laplace equation in two space dimensions and the
$\varphi _{1,2}$ solutions of equations 3 and 4 on the edge of the
Hall bar. The equations for $%
\psi _{1,2}$ are discredited by the Finite Element Method.\
Dirichlet boundary conditions are set at $50\;\mu m$ wide metal
contacts located at the left and right side of the bar for the local
case or $10\;\mu m$ wide contacts around $x=0$ at the top and bottom
edge of the bar for the non-local case. The generalized
Neumann boundary conditions (eq 2) are set everywhere else.
To solve the boundary value problem for a system of ordinary
differential equations 3 and 4\, we use a finite difference code that
implements the 3-stage Lobatto IIIa formula. The boundary conditions
are set to \ $\varphi _{1,2}=\psi _{1,2}$ inside the metal contacts.
Resistance for the non-local case is calculated as%
\begin{eqnarray}
R_{xx} =V\;I_{tot}^{-1},\;I_{tot}=I_{edge}+I_{bulk}, \nonumber \\[0.6\baselineskip]
V =\frac{1}{2}\left( \varphi _{11}-\varphi _{11^{\prime }}+\varphi
_{21}-\varphi _{21^{\prime }}\right) ,\nonumber \\[0.6\baselineskip]
I_{edge} =\varphi _{1}-\varphi _{2}+\varphi _{2^{\prime }}-\varphi
_{1^{\prime }}, \nonumber \\[0.6\baselineskip]
I_{bulk} =\sigma _{xy}^{\left( 1\right) }\left( \psi _{1}-\psi
_{1^{\prime
}}\right) -\sigma _{xx}^{\left( 1\right) }\int \frac{\partial \psi _{1}}{%
\partial y}dx+  \nonumber \\[0.6\baselineskip]
\sigma _{xy}^{\left( 2\right) }\left( \psi _{2}-\psi _{2^{\prime
}}\right) -\sigma _{xx}^{\left( 2\right) }\int \frac{\partial \psi
_{2}}{\partial y}dx, \label{Rxx}
\end{eqnarray}
where $V$ is the non-local potential difference, $I_{tot}$ is the total current flowing between contacts,$I_{edge}$ is the current flowing on the edge,$I_{bulk}$ is the current flowing through the 2D layer, $\varphi _{i1}$ and $\varphi _{i1^{\prime }}$ are
the potentials at the probe locations on the top and bottom edge correspondingly, $%
\varphi _{i}$ , $\psi _{i}$, and $\varphi _{i^{\prime }}$ , $\psi
_{i^{\prime }}$\ are the potentials at the opposite edges.
\begin{figure}[ht]
\includegraphics[width=8cm]{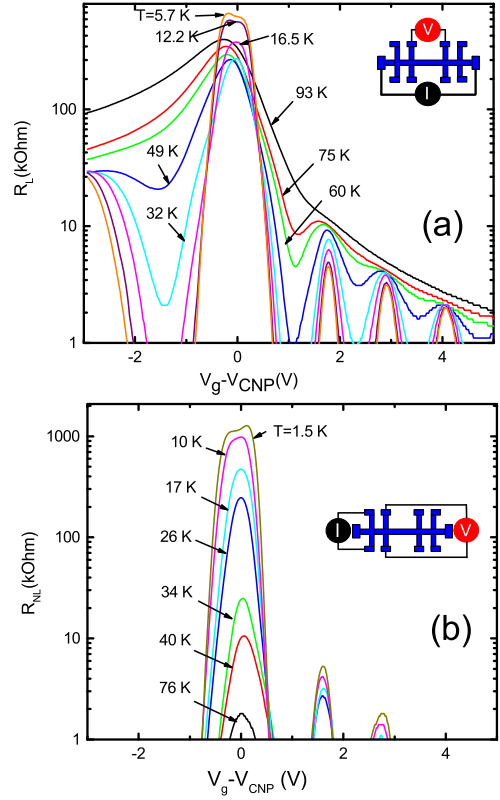}
 \caption{\label{fig.3}(Color online) The local $R_{L}=R_{xx}$ (I=1,6, V=3,4) (a)
and nonlocal $R_{NL}$ (I=2,10; V=3,9) (b) resistances as a function of the gate voltage and  temperature, B=5 T.
The schematics show how the current source and the voltmeter are connected for the
measurements.}
\end{figure}
For the local resistance configuration, all the potentials are calculated
in a similar way, but the variables for derivative and integration in
the equation for $I_{bulk}$ are interchanged. These give results entirely in
agreement with those found from the analytical solutions (Ref.12).
In summary, we provide the details of our modeling for the local and
non-local resistance as a function of carrier density $N_{s}$, which
we used to produce Fig. 4 in the main text. We focus on the QHE
regime (B =5 T; T =4.2 K). The best agreement between the experiment
and theory is reached for a value of the Landau level broadening for
electron-like and hole-like states A=13 , $\gamma^{-1}=0.33 \mu$m and
$g^{-1}=0.33\mu$m.  The magnetic field dependence of the nonlocal resistance has been modeled
from variation of $\Gamma$ (or A) with B, taken into account that the Zeeman splitting between LL peaks proportional to B.
The modeled behavior of the local and non-local resistivity is
consistent with our experimental observations. Note that the model we consider is fairly phenomenological,
and it is only valid at very low,
ideally zero temperature. Note, however, that model reproduced many observed features, in particular, peak values,
magnetic field and density dependencies.

\subsection{Temperature dependence of the local and nonlocal resistances in the presence of the magnetic field}

\begin{figure}[ht]
\includegraphics[width=10cm]{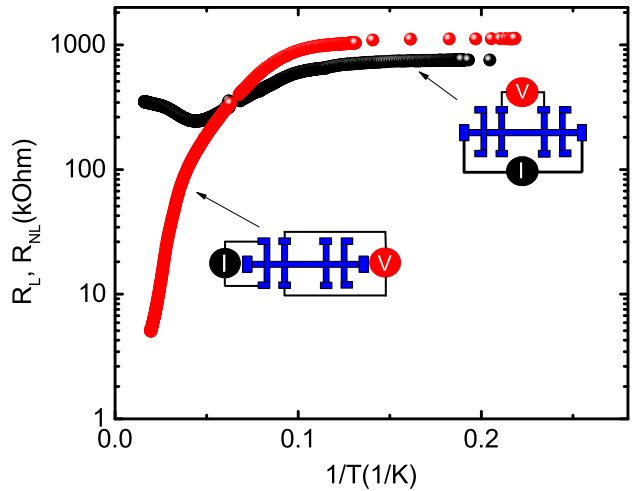}
 \caption{\label{fig.4}(Color online) The local $R_{L}=R_{xx}$ (I=1,6, V=3,4)
and nonlocal $R_{NL}$ (I=2,10; V=3,9) resistances at CNP as a function of the inverse temperature, B=5 T.
The schematics show how the current source and the voltmeter are connected for the
measurements.}
\end{figure}
In order to better understand the origin of the magnetic field induced nonlocality near CNP we studied the temperature dependence of
$R_{NL}$ and $R_{L}$. Figure 7 shows the evolution of local (a) and nonlocal (b) resistances voltage profiles
with temperature. There is  obvious difference in the  behaviour of the  two signals:
the peak in $R_{NL}$ at CNP dramatically decreases  above $T > 10 K$, while local resistance peak
is saturated at $T\approx 30 K$ and starts to rise at the higher temperature.
The different temperature regimes are more clearly revealed in the Figure 8. At low T
both peaks in $R_{NL}$ and $R_{L}$ decrease slowly with increasing T, whereas above $ \sim 10 K$,
one can see exponential decrease in both signals. Above $T > \sim 30 K$ peak in $R_{NL}$ follows a continuous
exponential decay, while local signal changes from decreasing to increasing.

The are number of physical mechanisms that can be responsible for the temperature dependence of the local and nonlocal
resistance at the CNP.  First, the nonlocal
resistance can be suppressed, because the charge carriers from localized states
 in Landau level tails are thermally excited into delocalized
states and contribute to conductivity, which shunts
the edge state transport at high temperature. Second, parameters $\gamma$ and
$g$ are expected be temperature dependent.
In general, determination of local and nonlocal resistances  requires a detailed knowledge of the distribution functions
at finite temperature,
found from a system of kinetic equations including both elastic and inelastic scattering. In the presence of edge to
bulk coupling, this problem is very complicated and out of scope of this paper.
In the simplified case, however, we could estimate that the nonlocal resistance drops in 3 order of
magnitude, assuming that the LL broadening increases in 1.3-1.5 times at highest temperature $(\sim 70 K)$,
which reasonably agrees with our observations (Figs 7 and 8).

\end{document}